\documentclass[aps,prd,preprint,superscriptaddress]{revtex4}
\usepackage{graphicx}

\begin{document}
\newcommand{\vecvar}[1]{\mbox{\boldmath$#1$}}

\preprint{OU-HET-619}

\title{Chiral-odd generalized parton distributions,\\
transversity decomposition of angular momentum, \\
and tensor charges of the nucleon}


\author{M.~Wakamatsu}
\affiliation{Department of Physics, Faculty of Science, \\
Osaka University, \\
Toyonaka, Osaka 560-0043, JAPAN}



\begin{abstract}
The forward limit of the chiral-odd generalized parton distributions
(GPDs) and their lower moments are investigated within the framework
of the chiral quark soliton model (CQSM), with particular emphasis
upon the transversity decomposition of nucleon angular momentum
proposed by Burkardt. A strong correlation between quark spin and
orbital angular momentum inside the nucleon is manifest itself in
the derived second moment sum rule within the CQSM, thereby
providing with an additional support to the qualitative connection
between chiral-odd GPDs and the Boer-Mulders effects.
We further confirm isoscalar
dominance of the corresponding first moment sum rule,
which indicates that the Boer-Mulders functions for the
$u$- and $d$-quarks have roughly equal magnitude with the
same sign. Also made are some comments on the recent empirical
extraction of the tensor charges of the nucleon by Anselmino et al.
We demonstrate that a comparison of their result with any
theoretical predictions must be done with great care, in consideration
of fairly strong scale dependence of tensor charges, especially at
lower renormalization scale.
\end{abstract}


\maketitle


\section{Introduction}

The concept of generalized parton distributions (GPDs) has
recently attracted considerable interest \cite{MRGDH94}
\nocite{DMRGH88}\nocite{Ji98}\nocite{GPV01}\nocite{Diehl03}-
\cite{BR05}. Naturally, these new quantities contain richer
information on the
internal quark-gluon structure of the nucleon, well beyond what
can be learned from the usual parton distribution functions (PDFs).
A complete set of quark GPDs at the leading twist 2 contains
four helicity conserving distributions, usually denoted as
$H^q, E^q, \tilde{H}^q, \tilde{E}^q$, and four helicity-flip
(chiral-odd) distributions, labeled as $H_T^q, E_T^q, \tilde{H}_T^q,
\tilde{E}_T^q$ \cite{HJ98},\cite{Diehl01}.
These GPDs are all functions of three kinematical
variables, $x$, $\xi$, and $t$, where $x$ is a generalized Bjorken
variable, $t$ is the four-momentum-transfer square of the nucleon,
while $\xi$ is the longitudinal momentum transfer, usually called
the skewdness parameter. The standard PFDs are naturally contained
as a subset of these GPDs. That is, in the forward limit
$\xi, \,t \rightarrow 0$ of zero momentum transfer,
$H^q (x,\xi,t)$ and $\tilde{H}^q (x,\xi,t)$ reduce to the unpolarized
distribution function $q(x)$, and the longitudinally polarized
distribution functions $\Delta q(x)$, respectively.
On the other hand, $H_T^q (x,\xi,t)$ reduces to the so-called
transversity distribution function $\Delta_T q(x)$.

Experimental studies so far have mostly been concentrated on the
helicity-conserving (chiral-even) GPDs, especially on
$H^q (x,\xi,t)$ and $E^q (x,\xi,t)$ \cite{ENV06}\nocite{Ellinghaus07}
\nocite{Ye06}-\cite{JLabHallA07}, because they are very interesting
quantities for clarifying the role of quark orbital angular
momentum in the nucleon spin problem \cite{Ji98},\cite{Ji97}
\nocite{HJL99}-\cite{JTH96} and also because they are
easier to access experimentally as compared with the helicity-flip
(chiral-odd) GPDs.
Although there exist some proposals to access to the chiral-odd GPDs
in diffractive double meson production \cite{IPST02},\cite{IPST04},
we now have almost no
empirical information on them. (An exception is the forward limit of
GPD $H^q (x,\xi,t)$, i.e. the transversity $\Delta_T q(x)$.
The first empirical extraction of the transversity distribution has
recently been done by Anselmino et al. based on the combined global
analysis of the measured azimuthal asymmetries in semi-inclusive
scatterings and those in $e^+ e^- \rightarrow h_1 h_2 X$
processes \cite{ABDKMPT07},\cite{ABDLMM08}.)

Although a direct experimental access to the chiral-odd GPDs is
not very easy at the present moment, it was shown by
Burkardt that they are not only interesting from a theoretical
viewpoint but also they have important influence on some
physical observables \cite{Burkardt05},\cite{Burkardt06}.
First, the transversity decomposition of quark
angular momentum in the nucleon introduced by him indicates that
a strong correlation between quark spin and angular momentum
is hidden in the 2nd moment of chiral-odd GPDs, more specifically
in the combination $H_T^q + 2 \,\tilde{H}_T^q + \tilde{E}_T^q$.
He also suggested that a strong correlation would exist between
the 1st moment of $2 \,\tilde{H}^q + \tilde{E}^q$ and the
Boer-Mulders functions $h_1^{\perp q}$ describing the asymmetry
of the transverse momentum of quarks perpendicular to the quark
spin in an unpolarized target \cite{BM98}.
(This is a variant of the analogous
relation between the Sivers function \cite{Sivers91} and the
anomalous magnetic moment of a quark with the flavor $q$ also
proposed by him \cite{Burkardt04A},\cite{Burkardt04B}.)

Turning to the status of theoretical studies of chiral-odd GPDs,
most works so far have been restricted to the studies of the
transversity $\Delta q(x)$, which is the forward limit of the
GPD $H_T^q (x,\xi,t)$. 
A lot of model calculations were reported on the transversity
and its 1st moment, i.e. the tensor charge \cite{Olness93}
\nocite{HJ95}\nocite{KPG96}\nocite{HJ96}\nocite{SS97}
\nocite{GRW98}\nocite{WK99}\nocite{SUPWPG01}\nocite{Wakamatsu01}
\nocite{ETZ04}\nocite{Wakamatsu07}-\cite{CBT08}.
There also exist lattice QCD studies on the lower moments of
$H_T^q (x,\xi,t)$. Its 1st moment, i.e. the tensor charge, was first
investigated in \cite{ADHK97}, while the simulations were extended to
include its 2nd moment as well in \cite{GHHPRSSZ05}.
However, the lattice QCD
studies on the moments of other GPDs, i.e.
$E_T^q, \tilde{H}_T^q, \tilde{E}_T^q$, have not been reported yet.
Concerning the full $x$, $\xi$, $t$ dependence of the chiral-odd
GPDs, there have been only a few model calculations.
The one is the investigation by Pasquini, Pincetti, and 
Boffi \cite{PPB05}\nocite{PPB06}-\cite{BP08} within the framework
of the light-front constituent quark model (see also a similar
investigation by Dahiya and Mukherjee \cite{DM08}), and
the other is the calculation by Scopetta based on a simple version
of the MIT bag model. Probably, most extensive is the investigation
by Pasquini et al. \cite{PPB05}.
They gave predictions not only for the lower
moments of GPDs but also for the full $x, \xi, t$-dependence of
those GPDs. Note however that their calculations were
made possible in price of one crude approximation. That is, in their
model calculations, only the lowest-order Fock-space components of
the light-front wave functions with three valence quarks are
taken into account.
The previous investigation of the chiral-even unpolarized GPD
$H^{(I=0)} (x,\xi,t)$ based on the approximate treatment of the
CQSM \cite{PPPBGW98} indicates that this approximation is
not necessarily justified,
and the inclusion of higher Fock-components may bring about richer
$x$- and $\xi$-dependence in the GPDs.

In the present investigation, we try to investigate chiral-odd
GPDs beyond the three valence quark approximation.
Within the CQSM, which we shall use, the effects of higher
Fock-space components can be included nonperturbatively as
contributions of deformed Dirac-sea quarks in the hedgehog mean
field \cite{DPP88},\cite{WY91},
not through the perturbative Fock-space expansion.
Unfortunately, technical hardness of this ambitious program
does not allow us complete calculation of GPDs in full dependence
of the three kinematical variables, $x, \xi$ and $t$, at the
present stage. In the present investigation, we therefore
content ourselves in the calculation of the forward limit of
a GPD, i.e. $G_T^q (x,0,0) \equiv \lim_{\xi \rightarrow 0, 
t \rightarrow 0} \,[\,H_T^q (x,\xi,t) + 2 \,\tilde{H}_T^q (x,\xi,t)
+ E_T^q (x,\xi,t) \,]$, since, except for the transversity distribution
$H_T^q (x,0,0)$, this distribution is physically the most interesting
quantity, which is thought to contain valuable information on
the correlation between the quark spin and orbital angular momentum
inside the nucleon as pointed out by Burkardt \cite{Burkardt05},
\cite{Burkardt06}. (There already exist several investigations
within the CQSM on the forward limit of chiral-even GPD
$E(x,\xi,t)$ \cite{OPSUG05},\cite{WT05} as well as on the
generalized form factors corresponding to its lower
moments \cite{WN06}\nocite{WN08}\nocite{GGOPSSU07A}-\cite{GGOPSSU07B}.)
Concerning the transversity $\Delta_T q(x) = H_T^q (x,0,0)$
contained in the combination to define $G_T^q (x,0,0)$,
the first empirical information has recently obtained by
Anselmino et al. through the semi-inclusive deep inelastic
scatterings \cite{ABDKMPT07},\cite{ABDLMM08}.
The extracted transversities and/or their 1st moments were compared
with some model predictions. We shall discuss in the present paper,
that such a comparison is potentially very dangerous if one does not
pay the closest attention to the fairly strong scale dependence of
the transversities.

The paper is organized as follows. First, in sect.\ref{Sect:formalism},
we shall derive theoretical formulas, which are
necessary for evaluating the relevant GPDs within the framework
of the CQSM. Next, in the first part of sect.\ref{Sect:results},
we show the results of numerical calculation for the isoscalar
and isovector part of $G_T^q (x,0,0)$ as well as their 1st and
2nd moments.    
The second part of sect.\ref{Sect:results} is devoted to
the discussion on the delicacy, which is shown to arise in
a comparison between the
recent empirical determination of the tensor charges and the
corresponding theoretical predictions.
Some concluding remarks are then given
in sect.\ref{Sect:conclusion}.

\section{Chiral-odd GPDs in the CQSM \label{Sect:formalism}}

The chiral-odd GPDs are defined as nonforward matrix elements of
light-cone correlation of the tensor current as
\begin{eqnarray}
 {\cal M} &=& 
 \frac{1}{2} \,\int \,\frac{d z^-}{2 \,\pi} \,e^{i \,x \,P^+ \,z^-} \,
 \langle p^\prime, \lambda^\prime \,| \,
 \bar{\psi} \left( - \,\frac{z}{2} \right) \,i \,
 \sigma^{+ j} \,\gamma_5 \,
 \psi \left( - \,\frac{z}{2} \right) \,|\, p, \lambda \rangle \nonumber \\
 &=& \frac{1}{2 \,P^+} \,
 \bar{u} (p^\prime, \lambda^\prime) \,\left[\,
 H_T (x,\xi,t) \,i \,\sigma^{+j} \,\gamma_5 \ + \ 
 \tilde{H}_T (x,\xi,t) \,
 \frac{i \,\epsilon^{+ j \alpha \beta} \,\Delta_\alpha \,P_\beta}{M_N^2}
 \right. \nonumber \\
 &\,& \hspace{20mm} \left. + \ E_T (x,\xi,t) \,
 \frac{i \,\epsilon^{+ j \alpha \beta} \,\Delta_\alpha \,\gamma_5}{2 \,M_N}
 \ + \ \tilde{E}_T (x,\xi,t)\,
 \frac{i \,\epsilon^{+ j \alpha \beta} \,P_\alpha \,\gamma_\beta}{M_N} \,
 \right] \,u (p, \lambda) , \label{Eq:ChiraloddGPDdef}
\end{eqnarray}
where $i = 1,2$ is a transverse index, while $p \,(p^\prime)$ and 
$\lambda \,(\lambda^\prime)$ are the momentum and the helicity of the
initial (final) nucleon, respectively.
We use here the light-cone coordinates $v^{\pm} = (v^0 \pm v^3) \,/ \,
\sqrt{2}$, and $\mbox{\boldmath $v$}_\perp = (v^1, v^2)$ for any
four-vector $v^\mu$. We also use the notation
\begin{equation}
 P \ = \ \frac{1}{2} \,(p^\prime \ + \ p), \ \ \ \ \ 
 \Delta \ = \ p^\prime - p.
\end{equation}
As is widely known, the GPDs depends on three kinematical variables,
$x$, $\xi$, and $t$, where $x$ is a generalized Bjorken variable,
$t = \Delta^2$ is the four-momentum-transfer squared of the nucleon,
and $\xi = - \,\Delta^+ \,/ \,(2 \,P^+)$ denotes the longitudinal
momentum transfer, usually called the skewdness parameter.

For model calculation, it is convenient to work in the so-called
Breit frame, in which
\begin{equation}
 p^\prime \ = \ \left( E_{\mbox{\boldmath $\Delta$} / 2}, 
 + \,\mbox{\boldmath $\Delta$} / 2 \right), \ \ \ \ \ 
 p \ = \ \left( E_{\mbox{\boldmath $\Delta$} / 2}, 
 - \,\mbox{\boldmath $\Delta$} / 2 \right), \ \ \ \ \ 
\end{equation}
so that
\begin{eqnarray}
 P &=& 
 \left( E_{\mbox{\boldmath $\Delta$} / 2}, \mbox{\boldmath $0$} \right),
 \ \ \ \ \ 
 \Delta \ = \ 
 \left( 0, \mbox{\boldmath $\Delta$} \right) .
\end{eqnarray}
We also assume large $N_c$ kinematics, in which the nucleon is heavy,
$M_N \sim O (N_c)$, and its center-of-mass motion is essentially
nonrelativistic.
Under these circumstances, $\Delta^i = O (N_c^0)$ and
$\Delta^0 = O (N_c^{-1})$, so that the hierarchy holds that
$M_N \gg | \Delta^i | \gg | \Delta^0 |$. Note also that
$t = - \,\mbox{\boldmath $\Delta$} = O (N_c^0)$ and 
$\xi = - \,\Delta^3 / (2 \,M_N) = O (N_c^{-1})$.
Then, noting that $\mbox{\boldmath $\Delta$} = 
(\mbox{\boldmath $\Delta$}_\perp , - \,2 \,M_N \,\xi)$, we evaluate the
right hand side of eq. (\ref{Eq:ChiraloddGPDdef}), to obtain
\begin{eqnarray}
 {\cal M} &\sim& \left[\,
 H_T \ + \ \frac{\mbox{\boldmath $\Delta$}_\perp^2}{8 \,M_N^2} \,
 \left( E_T - \frac{1}{2} \,H_T \right) \ + \ \xi \,\tilde{E}_T
 \,\right] \,\sigma_1 \nonumber \\
 &+& \frac{1}{2 \,M_N} \,\left[\, H_T \ + \ 2 \,\tilde{H}_T \ + \ E_T
 \,\right] \,i \,\Delta_2 \nonumber \\
 &+& \frac{1}{2 \,M_N} \,\tilde{E}_T \,\,\Delta_1 \,\sigma_3 \nonumber \\
 &-& \left[\,E_T \ - \ \frac{1}{2} \,H_T \,\right] \,
 \frac{1}{4 \,M_N^2} \,\left\{\, \Delta_1 \,
 (\mbox{\boldmath $\sigma$} \cdot \mbox{\boldmath $\Delta$}_\perp)
 \ - \ \frac{1}{2} \,\mbox{\boldmath $\Delta$}_\perp^2 \,\sigma_1
 \,\right\} .
\end{eqnarray}
Since we are interested in the forward limit
($\mbox{\boldmath $\Delta$}_\perp \rightarrow 0$, $\xi \rightarrow 0$)
in the present investigation, we can project out these four
independent pieces as
\begin{eqnarray}
 H_T (x,0,0) &=& \frac{1}{2 \,\pi} \,\int_0^{2 \,\pi} \,d \phi \,\,\,
 \frac{1}{2} \,\mbox{tr} \,\sigma_1 \,{\cal M}, \\
 \frac{1}{2 \,M_N} \,\left[\, H_T \ + \ 2 \,\tilde{H}_T \ + \ E_T \,\right]
 \,(x,0,0) &=& \frac{1}{\pi} \,\int_0^{2 \,\pi} \,d \phi \,\,
 \frac{\Delta_2}{i \,|\mbox{\boldmath $\Delta$}_\perp|^2} \,\,
 \frac{1}{2} \,\mbox{tr} \,{\cal M}, \label{Eq:Projection1} \\
 \frac{1}{2 \,M_N} \,\tilde{E}_T (x,0,0) &=& \frac{1}{\pi} \,
 \int_0^{2 \,\pi} \,\,d \phi \,\,
 \frac{\Delta_1}{|\mbox{\boldmath $\Delta$}_\perp|^2} \,\,
 \frac{1}{2} \,\mbox{tr} \,\sigma_3 \,{\cal M}, \label{Eq:Projection2} \\
 - \,\frac{1}{4 \,M_N^2} \,\left[\, E_T \ - \ \frac{1}{2} \,H_T \,\right]
 \,(x,0,0) &=& \frac{2}{\pi} \,\int_0^{2 \,\pi} \,d \phi \,\,
 \frac{1}{|\mbox{\boldmath $\Delta$}_\perp|^4} \nonumber 
 \label{Eq:Projection3}\\
 &\times& \frac{1}{2} \,\,
 \mbox{tr} \,\left[\,(\mbox{\boldmath $\Delta$}_\perp \cdot 
 \mbox{\boldmath $\sigma$} ) \,\Delta_1 \ - \ \frac{1}{2} \,
 \mbox{\boldmath $\Delta$}_\perp^2 \,\sigma_1 \,\right] \,{\cal M} ,
 \label{Eq:Projection4}
\end{eqnarray}
where $\phi$ is the azimuthal angle of the transverse vector
$\mbox{\boldmath $\Delta$}_\perp$, i.e. $\mbox{\boldmath $\Delta$}_\perp
= |\mbox{\boldmath $\Delta$}_\perp| \,(\cos \phi, \sin \phi)$.
For convenience, let us use below the shorthand notation : 
\begin{eqnarray}
 G_T (x,\xi,t) &\equiv&
 \ H_T (x,\xi,t) \ + \ 2 \,\tilde{H}_T (x,\xi,t)
 \ + \ E_T (x,\xi,t), \label{Eq:GPD_GT} \\
 K_T (x,\xi,t) &\equiv& 
 \ E_T (x,\xi,t) \ - \ \frac{1}{2} \,H_T (x,\xi,t).
\end{eqnarray}

Now, we are ready to evaluate the amplitudes ${\cal M}$ explicitly in the
CQSM. We first investigate the answer at the mean field level, i.e.
we derive theoretical expressions for
the $O (\Omega^0)$ contribution to ${\cal M}$, with $\Omega$ being the
collective angular velocity of the rotating hedgehog mean field.
Using the formalism developed in the previous studies \cite{WK99},
\cite{DPPPW96}\nocite{DPPPW97}\nocite{WGR96}\nocite{WGR97}
\nocite{WK98}\nocite{PPGWW99}\nocite{Wakamatsu03A}-\cite{Wakamatsu03B},
the isoscalar part of ${\cal M} (x,0,t)$ at the $O (\Omega^0)$
is given as a sum over
all the occupied eigen-states of the Dirac Hamiltonian $H$ :  
\begin{eqnarray}
 {\cal M}^{(I=0)} (x,0,t) &=& M_N \,\int \,\frac{d z_0}{2 \,\pi} \,
 \int \,d^3 \mbox{\boldmath $x$} \,\,
 e^{\,i \,\mbox{\boldmath $\Delta$}_\perp
 \cdot \mbox{\boldmath $x$}} \nonumber \\
 &\times& N_c \,\sum_{n \in occ} \,e^{i \,z_0 \,(x \,M_N - E_n)} \,
 \Phi_n^\dagger (\mbox{\boldmath $x$}) \,(\gamma_1 \,\gamma_5 - 
 i \,\gamma_2)
 \,\Phi_n (\mbox{\boldmath $x$} - z_0 \,\mbox{\boldmath $e$}_3) ,
\end{eqnarray}
with $\mbox{\boldmath $e$}_3 = (0,0,1)$ being a unit vector in the
$z$ direction. Here, $\Phi_n (\mbox{\boldmath $X$})$
are the eigenfunctions of the Dirac hamiltonian with the hedgehog mean
field, i.e.
\begin{equation}
 H \,\Phi_n (\mbox{\boldmath $x$}) \ = \ E_n \,
 \Phi_n (\mbox{\boldmath $x$}) ,
\end{equation}
with
\begin{equation}
 H \ = \ \frac{\mbox{\boldmath $\alpha$} \cdot \nabla}{i} \ + \ 
 \beta \,M \,e^{i \,\gamma_5 \,\mbox{\boldmath $\tau$} \cdot
 \hat{\mbox{\boldmath $r$}} \,F(r)} .
\end{equation}
Noting that
\begin{equation}
 e^{\,i \,\mbox{\boldmath $\Delta$}_\perp \cdot \mbox{\boldmath $x$}}
 \ = \ 
 1 \ + \ i \,\mbox{\boldmath $\Delta$}_\perp \cdot \mbox{\boldmath $x$}
 \ - \ \frac{1}{2} \,
 (\mbox{\boldmath $\Delta$}_\perp \cdot \mbox{\boldmath $x$})^2 \ + \ 
 \cdots , \label{Eq:Expansion}
\end{equation}
we easily find that
\begin{eqnarray}
 H_T^{(I=0)} (x,0,0) &=& 0 \\
 \frac{1}{2 \,M_N} \,G_T^{(I=0)} (x,0,0) &=&
 M_N \,N_c \,\sum_{n \in occ} \,\langle n \,| \,x_2 \,
 (\gamma_1 \,\gamma_5 - i \,\gamma_2) \,\delta (x \,M_N - E_n - p_3)
 \,| n \rangle, \hspace{10mm} \label{Eq:GTis} \\
 \frac{1}{2 \,M_N} \,\tilde{E}_T^{(I=0)} (x,0,0) &=& 0, \\
 \frac{1}{4 \,M_N^2} \,K_T^{(I=0)} (x,0,0) &=& 0 ,
\end{eqnarray}
which shows that only $G_T^{(I=0)} (x,0,0)$ survives among the
four GPDs, at the lowest order in $\Omega$, or in $1 / N_c$ expansion.

Next, we turn to the isovector part. The isovector part of ${\cal M}(x,0,t)$
is given as
\begin{eqnarray}
 {\cal M}^{(I=1)} (x,0,t) &=& M_N \,\int \,\frac{d z_0}{2 \,\pi} \,
 \int \,d^3 \mbox{\boldmath $x$} \,\,
 e^{\,i \,\mbox{\boldmath $\Delta$}_\perp
 \cdot \mbox{\boldmath $x$}} \nonumber \\
 &\times& N_c \,\sum_{n \in occ} \,e^{i \,z_0 \,(x \,M_N - E_n)} \,
 \Phi_n^\dagger (\mbox{\boldmath $x$}) \,
 A^\dagger \,\tau_3 \,A \,
 (\gamma_1 \,\gamma_5 - i \,\gamma_2)
 \,\Phi_n (\mbox{\boldmath $x$} - z_0 \,\mbox{\boldmath $e$}_3) , \ \ \ \ \ 
\end{eqnarray}
where $A$ is the rotation matrix belonging to flavor $SU(2)$.
Here we use the identity
\begin{equation}
 A^\dagger \,\tau_3 \,A \ = \ D_{3 k} (A) \,\tau_k ,
\end{equation}
where $D_{3 k} (A)$ is a Wigner's rotation matrix, which should
eventually be sandwiched between the rotational wave functions
$\Psi_{T_3,J_3}^{(T = S = 1/2)} [A]$ representing the spin-isospin states
of the nucleon. Using the expansion (\ref{Eq:Expansion}),
together with the replacement
\begin{equation}
 D_{3 k} (A) \ \longrightarrow \ - \,\frac{1}{3} \,
 \left( \tau_3 \right)_{T_3^\prime T_3} \,
 \left( \sigma_k \right)_{S_3^\prime S_3} ,
\end{equation}
which should be interpreted as an abbreviation of the identity
\begin{eqnarray}
 &\,& \int \,\Psi_{T_3^\prime, S_3^\prime}^{(T = S = 1/2)*} [A]
 \,\, D_{3 k} (A) \,\,
 \Psi_{T_3, S_3}^{(T = S = 1/2)} [A] \,\,{\cal D} A
 \ = \ - \,\frac{1}{3} \,
 \left( \tau_3 \right)_{T_3^\prime T_3} \,
 \left( \sigma_k \right)_{S_3^\prime S_3} ,
\end{eqnarray}
in the projection formulas (\ref{Eq:Projection1}) - 
(\ref{Eq:Projection4}), we are led to the following
expressions for the $O (\Omega^0)$ contributions to the isovector
parts of four GPDs as
\begin{eqnarray}
 H_T^{(I=1)} (x,0,0) &=& M_N \,\left(- \,\frac{N_c}{3} \right) \,
 \sum_{n \in occ} \,\langle n \,| \,\tau_1 \,
 (\gamma_1 \,\gamma_5 - i \,\gamma_2) \,
 \delta (x M_N - E_n - p_3) \,| \,n \rangle, \hspace{10mm} \\
 \frac{1}{2 \,M_N} \,G_T^{(I=1)} (x,0,0) &=& 0 , \\
 \frac{1}{2 \,M_N} \,\tilde{E}_T^{(I=1)} (x,0,0) &=& 
 M_N \,\left( - \,\frac{N_c}{3} \right) \,
 \sum_{n \in occ} \,\langle n \,| \,i \,x_1 \,\tau_3 \,
 (\gamma_1 \,\gamma_5 - i \,\gamma_2) \,
 \delta (x M_N - E_n - p_3) \,| \,n \rangle, 
 \hspace{10mm} \label{Eq:ETiv0} \\
 \frac{1}{4 \,M_N^2} \,K_T^{(I=1)} (x,0,0) &=& 
 M_N \,\left( - \,\frac{N_c}{3} \right) \,
 \sum_{n \in occ} \nonumber \\
 &\times& \langle n \,| \,\frac{1}{2} \,
 [\,2 \,(\mbox{\boldmath $x$}_\perp \cdot \mbox{\boldmath $\tau$}) \,x_1
 - \mbox{\boldmath $x$}_\perp^2 \,\tau_1 \,] \,
 (\gamma_1 \,\gamma_5 - i \,\gamma_2) \,
 \delta (x M_N - E_n - p_3) \,| \,n \rangle .
\end{eqnarray}
Note that, just opposite to the isoscalar case, only
$G_T^{(I=1)} (x,0,0)$ vanishes, while other three GPDs are
generally nonzero. 

The $O (\Omega^1)$ contributions, or equivalently, the next-to-leading
contributions in $1 / N_c$ expansion, can similarly be evaluated,
although the manipulation is much more complicated. Skipping the detailed
derivation, we first write down the answers for the isoscalar parts : 
\begin{eqnarray}
 &\,& H_T^{(I=0)} (x,0,0) \ = \ - \,M_N \,\frac{N_c}{2 \,I} \,
 \sum_{m \in all, n \in occ} \,\langle m \,|\,\tau_1 \,| \,n \rangle
 \nonumber \\
 &\,& \hspace{20mm}
 \times \ \langle n \,| \,(\gamma_1 \,\gamma_5 - i \,\gamma_2) \,
 \left( \frac{1}{E_m - E_n} - \frac{1}{2 \,M_N} \,\frac{d}{d x} \right) \,
 \delta (x \,M_N - E_n - p_3) \,| \, m \rangle , \\
 &\,& \frac{1}{2 \,M_N} \,G_T^{(I=0)} (x,0,0) \ = \ 0, \\
 &\,& \frac{1}{2 \,M_N} \,
 \tilde{E}_T^{(I=0)} (x,0,0) \ = \  - \,M_N \,\frac{N_c}{2 \,I} \,
 \sum_{m \in all, n \in occ} \,\langle m \,|\,\tau_3 \,| \,n \rangle
 \nonumber \label{Eq:ETis1} \\
 &\,& \hspace{15mm}
 \times \ \langle n \,| \,i \,x_1 \,
 (\gamma_1 \,\gamma_5 - i \,\gamma_2) \,
 \left( \frac{1}{E_m - E_n} - \frac{1}{2 \,M_N} \,\frac{d}{d x} \right) \,
 \delta (x \,M_N - E_n - p_3) \,| \, m \rangle , \hspace{15mm} \\
 &\,& \frac{1}{4 \,M_N^2} \,
 K_T^{(I=0)} (x,0,0) \ = \ - \,M_N \,\frac{N_c}{2 \,I} \,
 \sum_{m \in all, n \in occ} \,\langle m \,|\,\tau_c \,| \,n \rangle
 \nonumber \\
 &\,& \hspace{50mm}
 \times \ 
 \, \langle n \,| \,\frac{1}{2} \,[\,(x_1^2 - x_2^2) \,\delta_{c,1}
 \ + \ 2 \,x_1 \,x_2 \,\delta_{c,2} \,] \,
 (\gamma_1 \,\gamma_5 - i \,\gamma_2)
 \nonumber \\
 &\,& \hspace{50mm} \times \ 
 \left( \frac{1}{E_m - E_n} - \frac{1}{2 \,M_N} \,\frac{d}{d x} \right) \,
 \delta (x \,M_N - E_n - p_3) \,| \, m \rangle .
\end{eqnarray}
On the other hand, the $O (\Omega^1)$ contributions to the isovector part
are given as
\begin{eqnarray}
 &\,& H_T^{(I=1)} (x,0,0) \ = \ i \,\varepsilon_{1 a c} \,
 M_N \,\frac{N_c}{6 \,I} \,
 \sum_{m \in nocc, n \in occ} \,
 \frac{1}{E_m - E_n} \nonumber \\
 &\,& \hspace{50mm} \times \
 \langle m \,|\,\tau_c \,| \,n \rangle \,
 \langle n \,| \,\tau_a \,(\gamma_1 \,\gamma_5 - i \,\gamma_2) \,
 \delta (x \,M_N - E_n - p_3) \,| \, m \rangle , \\
 &\,& \frac{1}{2 \,M_N} \,G_T^{(I=1)} (x,0,0) \ = \ 
 - \,M_N \,\frac{N_c}{6 \,I} \,\sum_{m \in all, n \in occ} \,
 \,\langle m \,|\,\tau_c \,| \,n \rangle  \nonumber \\
 &\,& \hspace{15mm} \times \ 
 \langle n \,| \,\tau_c \,x_2 \,(\gamma_1 \,\gamma_5 - i \,\gamma_2) \,
 \left( \frac{1}{E_m - E_n} - \frac{1}{2 \,M_N} \,\frac{d}{d x} \right) \,
 \delta (x \,M_N - E_n - p_3) \,| \, m \rangle , \hspace{10mm} \\
 &\,& \frac{1}{2 \,M_N} \,
 \tilde{E}_T^{(I=1)} (x,0,0) \ = \ i \,\varepsilon_{3 a c} \,
 M_N \,\frac{N_c}{6 \,I} \,
 \sum_{m \in nocc, n \in occ} \,
 \frac{1}{E_m - E_n} \nonumber \\
 &\,& \hspace{40mm} \times \ 
 \,\langle m \,|\,\tau_c \,| \,n \rangle \,
 \langle n \,| \,\tau_a \,i \,x_1 \,
 (\gamma_1 \,\gamma_5 - i \,\gamma_2) \,
 \delta (x \,M_N - E_n - p_3) \,| \, m \rangle , 
 \hspace{10mm} \label{Eq:ETiv1} \\
 &\,& \frac{1}{4 \,M_N^2} \,
 K_T^{(I=1)} (x,0,0) \ = \ i \,\varepsilon_{b a c} \,
 M_N \,\frac{N_c}{6 \,I} \,
 \sum_{m \in nocc, n \in occ} \,\frac{1}{E_m - E_n} \,
 \langle m \,|\,\tau_c \,| \,n \rangle \nonumber \\
 &\,& \hspace{10mm} \times \ 
 \langle n \,| \,\tau_a \,\frac{1}{2} \,
 [\,(x_1^2 - x_2^2) \,\delta_{b,1}
 \ + \ 2 \,x_2 \,x_2 \,\delta_{b,2} \,] 
 \, (\gamma_1 \,\gamma_5 - i \,\gamma_2) \,
 \delta (x \,M_N - E_n - p_3) \,| \, m \rangle \,.
\end{eqnarray}
To sum up, we can summarize the novel $\Omega$ dependence of the
eight GPDs as
\begin{eqnarray}
 H_T^{(I=0)} (x,0,0) &=& \ \ 0 \ \ \ + \ O (\Omega^1),
 \label{Eq:countingrule1} \\
 G_T^{(I=0)} (x,0,0) &=& \ O (\Omega^0) \ + \ 0, \\
 \tilde{E}_T^{(I=0)} (x,0,0) &=& \ \ 0 \ \ \ + \ O (\Omega^1), \\
 K_T^{(I=0)} (x,0,0) &=& \ \ 0 \ \ \ + \ O (\Omega^1),
 \label{Eq:countingrule8}
\end{eqnarray}
and
\begin{eqnarray}
 H_T^{(I=1)} (x,0,0) &=& \ O (\Omega^0) \ + \ O (\Omega^1), 
 \label{Eq:IVHT} \\
 G_T^{(I=1)} (x,0,0) &=& \ \ \ 0 \ \ \ \ + \ O (\Omega^1), \\
 \tilde{E}_T^{(I=1)} (x,0,0) &=& \ O (\Omega^0) \ + \ O (\Omega^1),
 \label{Eq:IVGT} \\
 K_T^{(I=1)} (x,0,0) &=& \ O (\Omega^0) \ + \ O (\Omega^1) ,
 \label{Eq:IVKT}
\end{eqnarray}
where terms which do not contribute are denoted as $0$.
Since $\Omega$ is an $O (1 / N_c)$ quantity, this especially means
that $G_T (x,0,0)$ is a quantity with isoscalar dominance, although
the isovector component also exists as an $1 / N_c$ correction.
(See the discussion in sect.III.)
One may also notice that the contributions to the isovector GPDs,
$H_T^{(I=1)}$, $\tilde{E}_T^{(I=1)}$, and $K_T^{(I=1)}$,
survive at the mean-field level, or at the $O (\Omega^0)$ level,
while these GPDs receive $O (\Omega^1)$ contributions as well.
The appearance of the antisymmetric $\varepsilon$-tensor, as
observed in Eqs.(\ref{Eq:IVHT}), (\ref{Eq:IVGT}) and
(\ref{Eq:IVKT}), is a characteristic feature of
this novel $1 / N_c$ correction. This unique $1 / N_c$ correction
is known to play an important role for resolving the notorious
underestimation problem of some isovector observables of the nucleon
like the isovector axial charge and the isovector magnetic moment,
inherent in the hedgehog-type soliton model \cite{WW93},
\cite{CBGPPWW94},\cite{Wakamatsu96}. 

So far, we have derived the theoretical expressions for the forward
limits of four chiral-odd GPDs, $H_T (x,\xi,t)$, $E_T (x,\xi,t)$,
$\tilde{H}_T (x,\xi,t)$, and $\tilde{E}_T (x,\xi,t)$ in the CQSM.
Since the forward limit of $H_T (x,\xi,t)$, which is known to reduce
to the familiar transversity distribution $\Delta_T q(x)$, has
already been investigated within the CQSM \cite{WK99}\nocite{SUPWPG01}
\nocite{Wakamatsu01}-\cite{Wakamatsu07}, we need to evaluate
the remaining three independent GPDs, $E_T (x,0,0)$, $\tilde{H}_T (x,0,0)$,
and $\tilde{E}_T (x,0,0)$, or equivalently, $G_T (x,0,0)$, $K_T (x,0,0)$,
and $\tilde{E}_T (x,0,0)$. Unfortunately, we find that the numerical
calculation of $K_T (x,0,0)$ is quite involved. 
In the present study, we therefore concentrate on $G_T (x,0,0)$,
which is a special combination of three GPDs, $H_T (x,0,0)$,
$\tilde{H}_T (x,0,0)$, and $E_T (x,0,0)$, as given by (\ref{Eq:GPD_GT}).
From a physical viewpoint, this is the most interesting quantity,
which appears in Burkardt's transversity decomposition of quark
angular momentum \cite{Burkardt05},\cite{Burkardt06}.

According to Burkardt, the transverse decomposition of angular
momentum in the nucleon is given in the form
\begin{equation}
 \langle J_q^x \rangle \ = \ 
 \langle J_{q, + \hat{x}}^x \rangle \ + \ 
 \langle J_{q, - \hat{x}}^x \rangle ,
\end{equation}
where the 1st and the 2nd terms in the right hand side respectively
stand for the angular momentum carried by quarks with transverse
polarization in the $+ \,\hat{x}$ and $- \,\hat{x}$ directions in an
unpolarized nucleon at rest.
On the other hand, the difference of the above two quantities gives the
transverse asymmetry
\begin{equation}
 \langle \delta^x \,J_q^x \rangle \ = \ 
 \langle J_{q, + \hat{x}}^x \rangle \ - \ 
 \langle J_{q, - \hat{x}}^x \rangle ,
\end{equation}
which can be interpreted as representing a correlation between
quark spin and orbital angular momentum in an unpolarized nucleon.
Burkardt has derived the identities, which relate the above two
quantities to the 1st and the 2nd moment of GPDs as
\begin{eqnarray}
 \langle J_q^x \rangle &=& \frac{S^x}{2} \,\int_{-1}^1 \,
 \left[\, H (x,0,0) \ + \ E (x,0,0) \,\right] \,dx \\
 \langle \delta J_q^x \rangle &=& \frac{1}{2} \,\int_{-1}^1 \,x \,
 \left[\, H_T (x,0,0) \ + \ 2 \,\tilde{H}_T (x,0,0) \ + \ 
 E_T (x,0,0) \,\right] \, d x \nonumber \\
 &=& \frac{1}{2} \,\int_{=1}^1 \,x \,
 G_T (x,0,0) \,dx .
\end{eqnarray}
The quantities appearing in the 1st sum rule are the forward limit of
the familiar (chiral-even) unpolarized GPDs, $H (x,\xi,t)$ and
$E (x,\xi,t)$.
This identity is essentially Ji's nucleon spin sum rule and
nothing new \cite{Ji98}.
What is new is the second sum rule. It relates the above-mentioned
transverse asymmetry to the 2nd moment of the chiral-odd GPD $G_T$,
which we recall is a particular combination of $H_T$, $\tilde{H}_T$, and
$E_T$. It is interesting to see the explicit form for the 2nd
moment of $G_T^{(I=0)} (x,0,0)$ in the CQSM. From (\ref{Eq:GTis}),
we readily find that
\begin{eqnarray}
 &\,& \int_{-1}^1 \,x \,G_T^{(I=0)} (x,0,0) \,dx \nonumber \\
 &\,& = \,
 \frac{2}{3}\,N_c \,\sum_{n \in occ} \,\left\{ \,
 E_n \,\langle n \,| \,
 (- \,i) \,\mbox{\boldmath $\gamma$} \cdot \mbox{\boldmath $x$} \,
 | \,n \rangle \ + \ \frac{1}{2} \, \langle n \,|\,
 \gamma^0 \,\mbox{\boldmath $\Sigma$} \cdot \mbox{\boldmath $L$}
 \,| \,n \rangle \,\right\} \nonumber \\
 &\,& = \,
 \frac{2}{3}\,N_c \,\sum_{n \in occ} \,\left\{ \,
 E_n \,\langle n \,| \,
 \left( \begin{array}{cc}
 0 & \,- \,i \,\mbox{\boldmath $\sigma$} \cdot \mbox{\boldmath $x$} \\
 i \,\mbox{\boldmath $\sigma$} \cdot \mbox{\boldmath $x$} & 0 \\
 \end{array} \right)
 \,| \,n \rangle \ + \ \frac{1}{2} \, \langle n \,|\,
 \left( \begin{array}{cc}
 \mbox{\boldmath $\sigma$} \cdot \mbox{\boldmath $L$} & 0 \\
 0 & - \,\mbox{\boldmath $\sigma$} \cdot \mbox{\boldmath $L$} \\
 \end{array} \right)
 \,| \,n \rangle \,\right\} , \ \ \ \ \ \label{Eq:SR2nd} 
\end{eqnarray}
with $\mbox{\boldmath $\Sigma$} \equiv \gamma^0 \,\mbox{\boldmath $\gamma$}
\,\gamma_5$ being the relativistic spin operator of quark field.
As argued by Burkardt, the transverse asymmetry signals the correlation
between quark spin and orbital angular momentum in an unpolarized target.
One can clearly see that such correlation manifests itself in the
2nd term of the above sum rule (\ref{Eq:SR2nd}), since it reduces
to the nucleon matrix element (at the mean-field level) of the operator
$\gamma^0 \,\mbox{\boldmath $\Sigma$} \cdot \mbox{\boldmath $L$}$, which
is certainly the scalar product of the relativistic spin and orbital
angular momentum of quarks aside from an extra factor $\gamma^0$.
Unfortunately, the 1st term of eq.(\ref{Eq:SR2nd}) is a
highly model dependent expression, as convinced from the appearance
of the single-particle energy $E_n$ of the Dirac Hamiltonian $H$ with
the hedgehog mean field. This makes a simple physical interpretation of
the 1st term not so easy.

Before ending this section, we want to make a brief comment on
the GPD $\tilde{E}_T (x,\xi,t)$.
Although this GPD is generally nonzero 
(see Eqs. (\ref{Eq:ETiv0}), (\ref{Eq:ETiv1}), and (\ref{Eq:ETis1})),
its 1st moment is known to vanish by time reversal
invariance \cite{Diehl01}.
As a consistency check of our theoretical framework, we shall
explicitly prove in Appendix that the 1st moment of
$\tilde{E}_T (x,0,0)$ in fact vanishes identically.

\section{Numerical results and discussions \label{Sect:results}}

\subsection{chiral-odd GPDs and transversity decomposition of
angular momentum}

Within the framework of the CQSM, the expression of any nucleon
observable is divided into two parts, i.e. the contribution of
what-we-call the valence quark level (it is the lowest energy eigenstate
of a Dirac equation with the hedgehog mean field, which emerges from
the positive energy continuum) and that of the deformed Dirac sea
quarks. Since the latter contains ultraviolet divergences,
it must be regularized. Here, we use the Pauli-Villars regularization
scheme with single subtraction, for
simplicity \cite{DPPPW96}\nocite{DPPPW97}-\cite{WK99}.
The Pauli-Villars regulator mass
$M_{PV}$ is not an adjustable parameter of the model.
It is uniquely determined from a model consistency,
once the dynamical quark mass $M$, the only one parameter of the CQSM,
is fixed to be $M = 375 \,\mbox{MeV}$ from the phenomenology of the
nucleon low energy observables.

\begin{figure}[htb] \centering
\begin{center}
 \includegraphics[width=13.0cm,height=10.0cm]{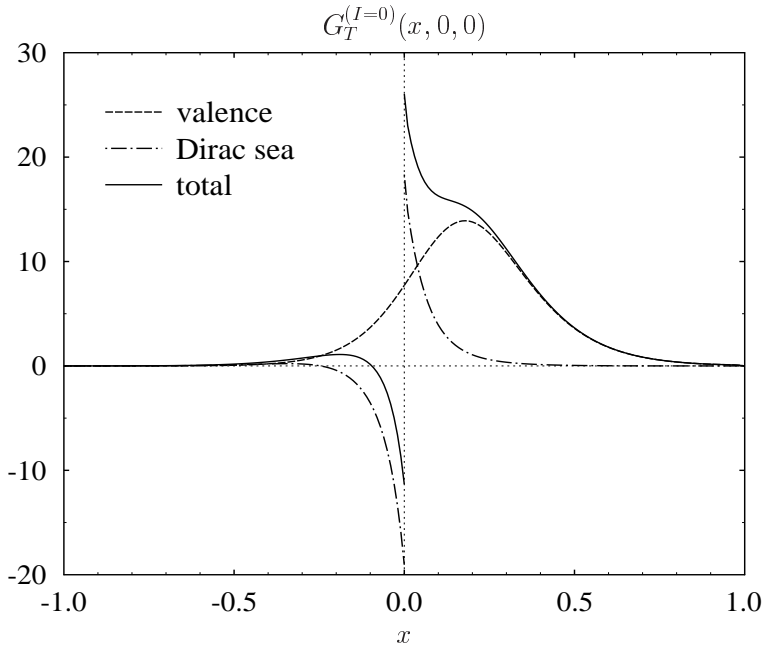}
\end{center}
\vspace*{-1.0cm}
\caption{The prediction of the CQSM for the forward limit of the
isoscalar GPD $G_T^{(I=0)}(x,0,0) \equiv \lim_{\xi \rightarrow 0,
t \rightarrow 0} \,[\,H_T^{(I=0)} (x,\xi,t) + 2 \,
\tilde{H}_T^{(I=0)} (x,\xi,t) + E_T^{(I=0)} (x,\xi,t) \,]$.
The dashed and dotted curves
respectively stand for the contributions of $N_c \,(\,= 3)$ valence
quarks and of deformed Dirac-sea quarks, while their sum is shown by
the solid curve.}
\label{Fig:GTis0}
\end{figure}%

We first show in Fig.\ref{Fig:GTis0} the CQSM predictions for the
forward limit of the isoscalar GPD $G_T^{(I=0)} (x,\xi,t)$, i.e.
$G_T^{(I=0)}(x,0,0) \ \equiv \ \lim_{\xi \rightarrow 0,
t \rightarrow 0} \,[\,H_T^{(I=0)} (x,\xi,t) \ + \ 2 \,
\tilde{H}_T^{(I=0)} (x,\xi,t) \ + \ E_T^{(I=0)} (x,\xi,t) \,]$.
Here, the dashed and dash-dotted curves respectively stand for the
contribution of $N_c \,(\,= 3\,)$ valence quarks and that of deformed
Dirac-sea quarks, while their sum is shown by
the solid curve. The distribution function in the negative
$x$ region should be interpreted as antiquark distribution except for
an extra minus sign related to the charge-conjugation property
of this distribution.
One clearly sees a strong chiral enhancement of
the deformed Dirac-sea contribution in the small $x$ region.
We recall the fact that a similar chiral enhancement
of the Dirac-sea contribution is also observed in the CQSM
prediction for more familiar unpolarized parton distribution function
of isoscalar type, and that it plays a crucial role for ensuring the
positivity condition of the antiquark distribution
$\bar{u}(x) + \bar{d}(x)$ \cite{DPPPW96},\cite{DPPPW97}.
Naturally, such chiral enhancement of the antiquark distribution
cannot be reproduced by such a model as light-cone constituent quark
model with $N_c \,(\,= 3)$ quark approximation.

\begin{figure}[htb] \centering
\begin{center}
 \includegraphics[width=13.0cm,height=10.0cm]{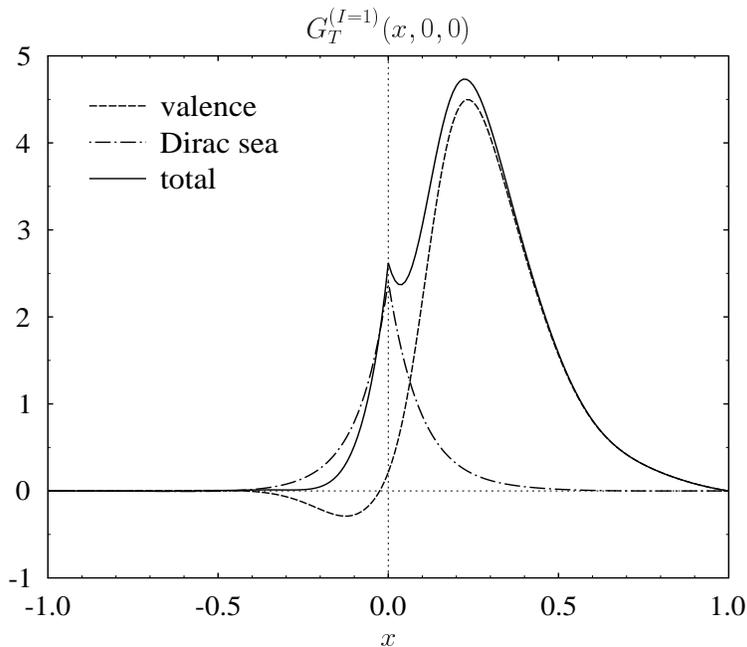}
\end{center}
\vspace*{-1.0cm}
\caption{The prediction of the CQSM for the forward limit of the
isovector GPD $G_T^{(I=1)}(x,0,0) \equiv \lim_{\xi \rightarrow 0,
t \rightarrow 0} \,[\,H_T^{(I=1)} (x,\xi,t) + 2 \,
\tilde{H}_T^{(I=1)} (x,\xi,t) + E_T^{(I=1)} (x,\xi,t) \,]$.
The meaning of the curves is the same as in Fig.\ref{Fig:GTis0}.}
\label{Fig:GTiv1}
\end{figure}

Next, shown in Fig.\ref{Fig:GTiv1} is the CQSM prediction for
the forward limit of the isovector GPD $G_T^{(I=1)} (x,\xi,t)$, i.e.
$G_T^{(I=1)}(x,0,0) \ \equiv \ \lim_{\xi \rightarrow 0,
t \rightarrow 0} \,[\,H_T^{(I=1)} (x,\xi,t) \ + \ 2 \,
\tilde{H}_T^{(I=1)} (x,\xi,t) \ + \ E_T^{(I=1)} (x,\xi,t) \,]$.
Here, the meaning of the curves is the same as in the previous figure.
Also for the isovector distribution, one observes a strong chiral
enhancement of the deformed Dirac-sea contribution in the small $x$ region.
The $x$ dependence of the Dirac-sea contribution for this isovector
distribution turns out to be totally different from the isoscalar
distribution, however.
The deformed Dirac-sea contribution for the isovector
distribution is nearly symmetric with respect to the variable change
$x \rightarrow - \,x$, in sharp contrast to the isoscalar
distribution, which is approximately antisymmetric.
This behavior is again resembling more familiar unpolarized parton
distribution function of isovector type \cite{WK98},\cite{PPGWW99},
\cite{Wakamatsu03A}.
We recall the fact that this chiral enhancement of the isovector
unpolarized distribution is just what is required by the celebrated
NMC measurement, which established the dominance of $\bar{d}$ sea
over the $\bar{u}$ sea inside the proton \cite{NMC91}.
Unfortunately, we do not
have any simple explanation about why such a similarity exists between 
the small $x$ behaviors of the unpolarized parton distribution functions
$q(x)$ and the forward limit of the GPD $G_T (x,\xi,t)$.

As discussed before, the above two distribution, i.e, 
$G_T^{(I=0)}(x,0,0)$ and $G_T^{(I=1)}(x,0,0)$, are very interesting quantities from a physical viewpoint, since they give the
transversity decomposition of the angular momentum inside
the nucleon. So far, there exist only a few theoretical investigations
on the chiral-odd GPDs and their moments.
In table \ref{Tab:transasym}, we compare the CQSM predictions
for the transverse asymmetries
with those of the two versions of the light-front constituent quark model
by Pasquini et al., that is the harmonic oscillator (HO) model and
the hyper central model \cite{PPB05}.
Note that their models are essentially the three quark model
with relativistic kinematics.
One finds that the predictions of the CQSM just lie between the HO model
and the hyper central model.
In all the three models, the isoscalar transverse asymmetry is seen
to be larger than the isovector one, but the isoscalar-to-isovector
ratio is largest in the CQSM.
This observation is qualitatively consistent with
the large $N_c$ prediction given in \cite{Pobylitsa05}.
However, literally taking the
$N_c \rightarrow \infty$ limit, the ratio
$\langle \delta^x J_x^{u+d} \rangle \,/ \,\langle 
\delta^x J_x^{u-d} \rangle$ would become infinite.
Our present analysis here shows
that $G_T (x,\xi,t)$ is an isoscalar-dominant quantity but the isovector
component, which arises as a $1 / N_c$ correction, is also important.

\vspace{4mm}
\begin{table}[h]
\begin{center}
\begin{tabular}{cccc} \hline
\ \ \ transverse asymmetry \ \ \ & \ \ HO \ \ & \ \ Hypercentral \ \ 
& \ \ CQSM \ \ \\ \hline
$\langle \delta^x J_x^u \rangle$ & 0.68 & 0.39 & 0.49 \\
$\langle \delta^x J_x^d \rangle$ & 0.28 & 0.10 & 0.22 \\ \hline
$2 \,\langle \delta^x J_x^{u+d} \rangle$ & 1.92 & 0.98 & 1.41 \\
$2 \,\langle \delta^x J_x^{u-d} \rangle$ & 0.80 & 0.58 & 0.54 \\ \hline
\ \ \ $\langle \delta^x J_x^{u+d} \rangle \,/ \,\langle 
\delta^x J_x^{u-d} \rangle$ \ \ \ & 2.40 & 1.69 & 2.61 \\ \hline
\end{tabular}
\vspace{3mm}
\caption{Some theoretical predictions for the transverse asymmetry.
Here, the second and the third columns stand for the two versions of
the light-front constituent quark model of Pasquini
et al. \cite{PPB05}, i.e.
the harmonic oscillator model (HO) and the hyper central model (HYP),
while the predictions of the CQSM are shown in the fourth column.}
\label{Tab:transasym}
\end{center}
\end{table}

Also interesting is the 1st moment sum rule for $G_T$, 
which can be divided into two pieces, i.e. the 1st moment
of the transversity $H_T^q (x,0,0)$ and that of the distribution
$2 \,\tilde{H}_T^q (x,0,0) + E_T^q (x,0,0)$ as
\begin{eqnarray}
 \int_{-1}^1 \,G_T^q (x,0,0) \,dx &=&
 \int_{-1}^1 \,H_T^q (x,0,0) \,dx \ + \ 
 \int_{-1}^1 \,[\,2 \,\tilde{H}_T^q (x,0,0) \ + \ 
 E_T^q (x,0,0) \,] \,dx . \label{Eq:decomposition}
\end{eqnarray}
Here, the 1st term of the r.h.s. of the above equation, i.e.
the 1st moment of the transversity, gives the tensor charge
\begin{equation}
 \Delta_T q \ = \ \int_{-1}^1 \,H_T^q (x,0,0) \,dx.
\end{equation}
On the other hand, the 2nd term, i.e. the 1st moment of 
$2 \,\tilde{H}_T^q (x,0,0) + E_T^q (x,0,0)$ defined by
\begin{equation}
 \kappa_T^q \ = \ \int_{-1}^1 \,
 [\,2 \,\tilde{H}_T^q (x,0,0) \ + \ 
 E_T^q (x,0,0) \,] \,dx ,
\end{equation}
was given an interpretation as a quantity governing the transverse
spin-flavor dipole moment in an unpolarized target by
Burkardt. In fact, he showed that $\kappa_T^q$ gives us an
information on how far and
in which direction the average position of quarks with spin in the
$\hat{x}$ direction for an unpolarized target relative to the center
of momentum.
The decomposition (\ref{Eq:decomposition}) corresponds to a
similar decomposition of the
1st moment sum rule for the unpolarized GPD,
$E_M^q (x,\xi,t) \equiv H^q (x,\xi,t) + E^q (x,\xi,t)$,
which gives the total magnetic moment consisting of
the quark number $N^q$ and the anomalous magnetic moment
$\kappa^q$ as
\begin{eqnarray}
 \int_{-1}^1 \,E_M^q (x,0,0) \,dx &=&
 \int_{-1}^1 \,H^q (x,0,0) \, dx \ + \ \int_{-1}^1 \,E^q (x,0,0) \,dx
 \ = \ 
 \ \ N^q \ + \ \kappa^q .
\end{eqnarray}
In table \ref{Tab:atm}, we again compare the CQSM predictions for
the quantities $\kappa_T^q$ (here we tentatively call it the
``anomalous tensor moment'')
with the corresponding predictions of Pasquini et al.
Here, the prediction for the isoscalar part is closer to that of
the HO model, while the prediction for the
isovector part is closer to that of the hyper central model.

\vspace{6mm}
\begin{table}[h]
\begin{center}
\begin{tabular}{cccc} \hline
\ \ \ 1st moment of $2 \,\tilde{H}_T + E_T$ 
\ \ \ & \ \ HO \ \ & \ \ Hypercentral \ \ 
& \ \ CQSM \ \ \\ \hline
$\kappa_T^u$ & 3.60 & 1.98 & 3.47 \\
$\kappa_T^d$ & 2.36 & 1.17 & 2.60 \\ \hline
$\kappa_T^{u+d}$ & 5.96 & 3.15 & 6.07 \\
$\kappa_T^{u-d}$ & 1.24 & 0.81 & 0.88 \\ \hline
\ \ \ $\kappa_T^{u+d} \,/ \,\kappa_T^{u-d}$ \ \ \ 
& 4.81 & 3.89 & 6.90 \\ \hline
\end{tabular}
\vspace{3mm}
\caption{The theoretical predictions for the 1st moment of
$2 \,\tilde{H}_T + E_T$.}
\label{Tab:atm}
\end{center}
\end{table}

According to Burkardt's conjecture, one would expect an intimate
connection between the time-reversal-odd (T-odd) transverse
momentum-dependent distributions and the GPDs.
They are the approximate proportionality relation between
Siver's function and the anomalous magnetic moment with opposite
sign \cite{Burkardt04A},\cite{Burkardt04B},
\begin{equation}
 f_1^{\perp q} (x,\mbox{\boldmath k}_\perp^2) \ \sim \ - \,
 \kappa^q ,
\end{equation}
and also the proportionality relation between Boer-Mulders' function
and the anomalous tensor moment \cite{Burkardt05},\cite{Burkardt06},
\begin{equation}
 h_1^{\perp q} (x,\mbox{\boldmath k}_\perp^2) \ \sim \ - \,
 \kappa_T^q .
\end{equation}
If his conjecture is combined with with some typical model
predictions for the anomalous tensor moments, one would get the
following approximate relations :
\begin{eqnarray*}
 h_1^{\perp d} \ \sim \ \frac{1}{2} \,h_1^{\perp u} \ \ &:& \ \ 
 \mbox{MIT bag model} , \\
 h_1^{\perp d} \ \sim \ h_1^{\perp u} \ \ \ \ &:& \ \ 
 \mbox{Large $N_c$ prediction}, \\ 
 h_1^{\perp d} \ \sim \ \frac{3}{4} \,h_1^{\perp u} \ \ &:& \ \ 
 \mbox{CQSM} ,
\end{eqnarray*}
thereby dictating that the Boer-Mulders functions for the $u$- and
$d$-quarks would have the same sign, although the predictions
on the relative magnitudes are a little variant.
This should be contrasted with the fact that Sivers functions for
the $u$- and $d$-quarks appears to have opposite sign as
\begin{equation}
 f_1^{\perp d} \ \sim \ - \,f_1^{\perp u} ,
\end{equation}
in conformity with empirically known relation
\begin{equation}
 \kappa^d \ \sim \ - \,\kappa^u .
\end{equation}
From our viewpoint, the origin of this qualitative difference is
very simple. It comes from the fact that the anomalous
magnetic moment is a quantity with isovector dominance,
whereas the quantities $\kappa_T^q$ is of isoscalar dominance,
as expected from the $N_c$ counting rule indicated in
Eqs.(\ref{Eq:countingrule1})-(\ref{Eq:countingrule8}).

\subsection{tensor charges : current empirical information versus theoretical predictions}

Some years ago, the first empirical extraction of the transversity
distributions has been made by Anselmino et al. based on the combined
global analysis of the measured azimuthal asymmetries in semi-inclusive
deep inelastic scatterings (SIDIS) and those in
$e^+ e^- \rightarrow h_1 h_2 X$ processes.
More recently, they have further refined their global analysis by
using new data from HERMES, COMPASS, and BELLE
Collaborations \cite{ABDLMM08}. The 1st $x$-moments of
the transversity distributions - related to the tensor charge - have
been extracted to be
\begin{equation}
 \Delta_T u \ = \ 0.59 \ 
 \begin{array}{l}
 \raisebox{-0.4ex}[0pt]{+ \,0.14} \\
 \raisebox{+0.4ex}[0pt]{\,-- \,\,0.13} \\
 \end{array}, \ \ \ \ \ 
 \Delta_T d \ = \ - \,0.20 \ 
 \begin{array}{l}
 \raisebox{-0.4ex}[0pt]{+ \,0.05} \\
 \raisebox{+0.4ex}[0pt]{\,-- \,\,0.07} \\
 \end{array},
\end{equation}
at the renormalization scale $Q^2 = 0.8 \,\mbox{GeV}^2$. 
They concluded that their new transversity distributions are close
to some model predictions, especially the predictions by a covariant
quark-diquark model by Clo\"{e}t at al. \cite{CBT08}.
This agreement is related to the fact that the predictions
by Clo\"{e}t et al. give the smallest magnitudes of tensor charges
among many theoretical predictions including those of the lattice QCD.
As we shall discuss below, this statement appears very misleading,
however. A delicate point is that the tensor charges
are strongly scale-dependent quantities especially at low
renormalization scale. In fact, the bare predictions
of the covariant quark-diquark model given in \cite{CBT08}
for the tensor charges are nothing small. They are
\begin{equation}
 \Delta_T u \ = \ 1.04, \ \ \ \ \ 
 \Delta_T d \ = \ - \,0.24 ,
\end{equation}
or in the isospin language, 
\begin{equation}
 \Delta_T \,q^{(I=1)} \ = \ 1.28, \ \ \ \ \ 
 \Delta_T \,q^{(I=0)} \ = \ 0.80 .
\end{equation}
Clo\"{e}t et al. regard the transversity distributions, which give
the above 1st moments, as initial distributions given at the scale
$Q^2 = 0.16 \,\mbox{GeV}^2$, and take account of their scale
dependencies by using the next-to-leading (NLO) evolution equation.
This procedure gives the tensor charges at the scale
$Q^2 = 0.4 \,\mbox{GeV}^2$ : 
\begin{equation}
 \Delta_T u \ = \ 0.69, \ \ \ \ \ 
 \Delta_T d \ = \ - \,0.16 ,
\end{equation}
or equivalently
\begin{equation}
 \Delta_T q^{(I=1)} \ = \ 0.85, \ \ \ \ \ 
 \Delta_T q^{(I=0)} \ = \ 0.53 ,
\end{equation}
which are much smaller than the bare predictions of the model
despite pretty small scale difference.

\begin{figure}[htb] \centering
\begin{center}
 \includegraphics[width=13.0cm,height=10.0cm]{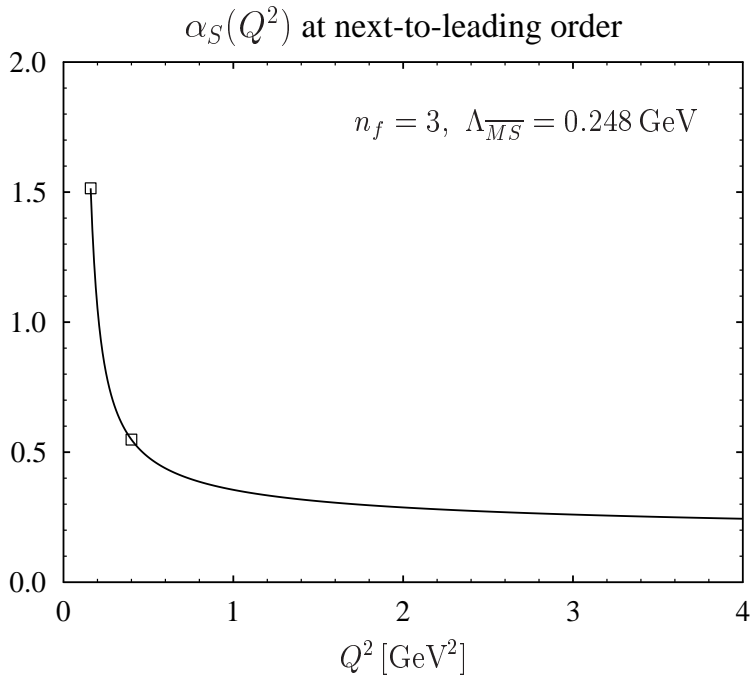}
\end{center}
\vspace*{-1.0cm}
\caption{The QCD running coupling constant $\alpha_S (Q^2)$ at the
NLO in dependence of $Q^2$, obtained with the effective flavor number
$n_f = 3$ and the QCD scale parameter
$\Lambda_{\overline{MS}} = 0.248 \,\mbox{GeV}$. The energy scale
$Q^2 = 0.16 \,\mbox{GeV}^2$ and $Q^2 = 0.40 \,\mbox{GeV}^2$ are
marked by open squares as a guide.}
\label{Fig:alphas}
\end{figure}%

As naturally anticipated, to start the NLO evolution at such low energy
scale as $Q^2 = 0.16 \,\mbox{GeV}^2$ is very dangerous.
To convince it more concretely, we first show in Fig.\ref{Fig:alphas}
the QCD running coupling constant $\alpha_S (Q^2)$ at the NLO as
a function of $Q^2$. Here, we have used the standard NLO formula
\begin{equation}
 \alpha_S^{NLO} (Q^2) \ = \ 
 \frac{4 \,\pi}{\beta_0 \,\ln (Q^2 / \Lambda^2)} \,
 \left[\,1 \ - \ \frac{\beta_1}{\beta_0^2} \,
 \frac{\ln \ln (Q^2 / \Lambda^2)}{\ln (Q^2 / \Lambda^2)} \,\right] ,
\end{equation}
where
\begin{equation}
 \beta_0 \ = \ 11 \ - \ \frac{2}{3} \,n_f, \ \ \ \ 
 \beta_1 \ = \ 102 \ - \ \frac{38}{3} \,n_f ,
\end{equation}
together with the effective flavor number $n_f = 3$ and the
QCD scale parameter
$\Lambda = \Lambda_{\overline{MS}} = 0.248 \,\mbox{GeV}$,
taken from the NLO analysis by Gl\"{u}ck, Reya, and Vogt \cite{GRV95}.
One sees that, at $Q^2 = 0.16 \,\mbox{GeV}^2$, $\alpha_S (Q^2)$ is
about 1.5, which immediately throw doubt on the use of the
perturbative QCD evolution equation.

The statement can be made more explicit by investigating the
NLO evolution of the tensor charges themselves.
The anomalous dimensions at the NLO, which control the scale
dependencies of the moments of the transversities are given in
\cite{KM97},\cite{HKK97},\cite{Vogelsang98}.
We are interested here in the NLO evolution of
the 1st moment, i.e. the tensor charges.
(Note that, since the transversities do not couple
to the gluon distributions, the evolution of the tensor charges is
flavor independent. For more detail, see the discussion later.)
The solution of the NLO evolution equation for the tensor charge
$\Delta_T q (Q^2)$ is given as
\begin{equation}
 \frac{\Delta_T \,q (Q^2)}{\Delta_T \,q (\mu^2)} \ = \ 
 \left( \frac{\alpha_S (Q^2)}{\alpha_S (\mu^2)} 
 \right)^{\gamma^{(0)} / 2 \beta_0} \,\,
 \left[\,\frac{\beta_0 \ + \ \beta_1 \,\alpha_S (Q^2) / 4 \,\pi}
 {\beta_0 \ + \ \beta_1 \,\alpha_S (\mu^2) / 4 \,\pi} \,
 \right]^{\frac{1}{2} \,\left(\,\gamma^{(1)} / \beta_1 \ - \  
 \gamma^{(0)} / \beta_0 \right)} , \label{Eq:exactRGE}
\end{equation}
with
\begin{eqnarray}
 \gamma^{(1)} / 2 \beta_1 &=& \left( \,\frac{724}{9} \ - \ 
 \frac{104}{27} \,n_f \,\right) \, / \,
 2 \,\left( \,102 \ - \ \frac{38}{3} \,n_f \,\right) , \\
 \gamma^{(0)} / 2 \beta_0 &=& 4 \,/ \,\left(\,33 - 2 \,n_f \,\right) .
\end{eqnarray}
To NLO accuracy, the above solution are sometimes expanded as
\begin{eqnarray}
 \frac{\Delta_T q (Q^2)}{\Delta_T q (\mu^2)} &=&
 \left( \frac{\alpha_S (Q^2)}{\alpha_S (\mu^2)} 
 \right)^{\gamma^{(0)} / 2 \beta_0} \,
 \left\{\,1 \ - \ \frac{1}{4 \,\pi} \,\frac{\beta_1}{\beta_0} \,
 \left(\, \frac{\gamma^{(1)}}{2 \,\beta_1} - 
 \frac{\gamma^{(0)}}{2 \,\beta_0} \,\right) \,
 [\,\alpha_S (\mu^2) \, - \, \alpha_S (Q^2) \,] \,\right\} 
 \ \ \ \nonumber \\
 &=& \left( \frac{\alpha_S (Q^2)}{\alpha_S (\mu^2)} 
 \right)^{4 / 27} \,\,
 \left\{\, 1 \ - \ \frac{337}{486 \,\pi} \,
 [\,\alpha_S (\mu^2) \, - \, \alpha_S (Q^2) \,] \,\right\} .
 \label{Eq:approximateRGE}
\end{eqnarray}
Here, we have set $n_f = 3$, which reproduces the form used
in \cite{CBT08}.
For large enough $Q^2$, where the QCD running coupling constant
is much smaller than unity, both expressions should approximately
be equivalent. However, we have already pointed out that, at
the scale of $Q^2 = 0.16 \,\mbox{GeV}^2$, $\alpha_S$ is even larger
than 1.5. 

\begin{figure}[htb] \centering
\begin{center}
 \includegraphics[width=13.0cm,height=10.0cm]{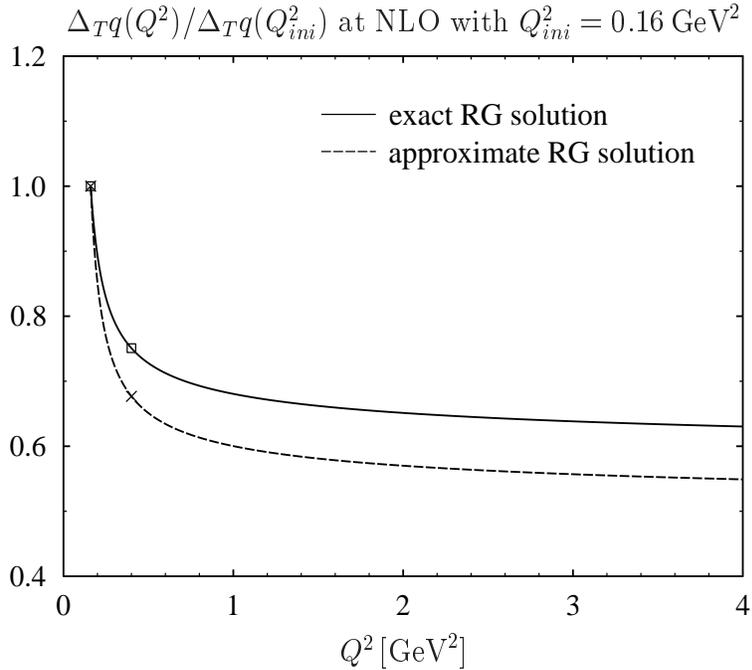}
\end{center}
\vspace*{-0.5cm}
\renewcommand{\baselinestretch}{1.20}
\caption{The scale dependence of the tensor charge, where the
evolution is started at $\mu^2 = Q_{ini}^2 = 0.16 \,\mbox{GeV}^2$.
The solid and dashed curves respectively correspond to the results
obtained with the exact (Eq.(\ref{Eq:exactRGE})) and approximate
(Eq.(\ref{Eq:approximateRGE}) solution of the NLO evolution equation.}
\label{Fig:tchg016}
\end{figure}%

Shown in Fig.\ref{Fig:tchg016} are the $Q^2$-dependence of tensor
charge, in which the evolution is started at
$\mu^2 = Q_{ini}^2 = 0.16 \,\mbox{GeV}^2$.
The solid and dashed curves respectively correspond to the
answers obtained by using the exact (Eq.(\ref{Eq:exactRGE})) and
approximate (Eq.(\ref{Eq:approximateRGE})) solutions of the
NLO evolution equation.
One clearly observes a drastic difference between the two choices.

\begin{figure}[htb] \centering
\begin{center}
 \includegraphics[width=13.0cm,height=10.0cm]{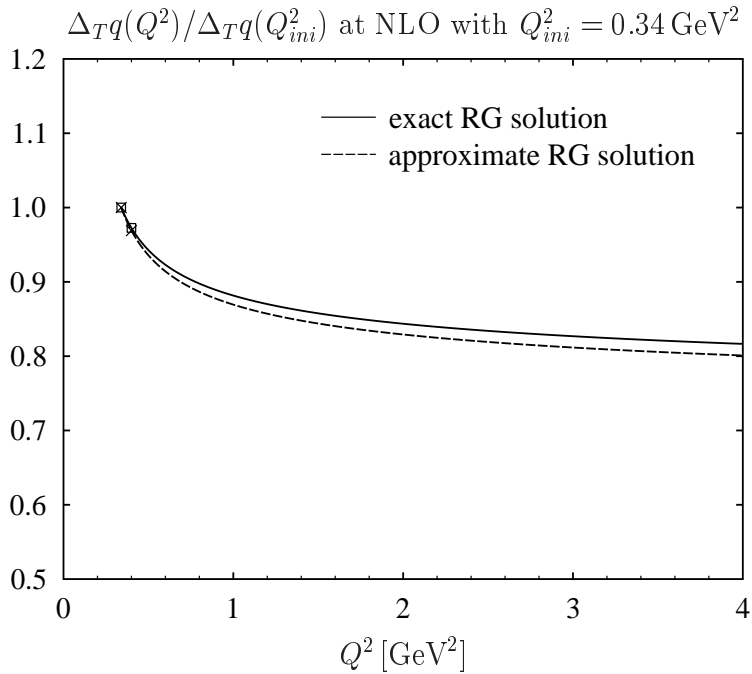}
\end{center}
\vspace*{-1.0cm}
\caption{The scale dependence of the tensor charge, where the
evolution is started at $\mu^2 = Q_{ini}^2 = 0.34 \,\mbox{GeV}^2$.
The solid and dashed curves respectively correspond to the results
obtained with the exact (Eq.(\ref{Eq:exactRGE})) and approximate
(Eq.(\ref{Eq:approximateRGE})) solution of the NLO evolution equation.}
\label{Fig:tchg034}
\end{figure}%

On the other hand, shown in Fig.\ref{Fig:tchg034} are the $Q^2$
dependence of the same quantity, where the evolution is
started at $\mu^2 = Q_{ini}^2 = 0.34 \,\mbox{GeV}^2$,
which corresponds to the choice adopted in the well-known NLO
analysis of the parton distribution functions by Gl\"{u}ck, Reya
and Vogt. The difference between the two
forms of NLO evolution solutions is fairly small, in this case.
One might suspect that not only the scale $Q_{ini}^2 = 0.16 \,
\mbox{GeV}^2$ but also the scale
$Q_{ini}^2 = 0.34 \,\mbox{GeV}^2$ is not high enough for the
perturbative QCD framework to be justified perfectly. 
This cannot be denied completely. Still, it is clear from our
simple analysis that there is a qualitative
difference between the two choices of the starting energy,
i.e. $Q_{ini}^2 = 0.16 \,\mbox{GeV}^2$ and
$Q_{ini}^2 = 0.34 \,\mbox{GeV}^2$.
As already pointed out, the authors of \cite{CBT08} use an
approximate solution of the NLO evolution equation with the choice
$Q_{ini}^2 = 0.16 \,\mbox{GeV}^2$ to estimate the tensor
charges at the scale $Q^2 = 0.40 \,\mbox{GeV}^2$.
The reduction of the magnitude of tensor charge after this
scale change is significant. It is about 0.75 if one uses
Eq.(\ref{Eq:exactRGE}), while it is about 0.67 if one uses
Eq.(\ref{Eq:approximateRGE}). (See the open squares and the
crosses in Fig.\ref{Fig:tchg016}.)
Undoubtedly, this enormous
reduction has nothing to do with the nature of their effective model.
It is simply a consequence of starting the NLO evolution
equation at such a low energy scale.

Generally, for any effective models of baryons, it is very
hard to say exactly what energy scale the predictions of
those correspond to. Probably, the best we can do at the moment
is to follow the spirit of PDF fit by Gl\"{u}ck et al. \cite{GRV95},
and use the predictions of those models as initial-scale
distributions given at the energy scale around
$600 \,\mbox{MeV}$, or $Q_{ini}^2 \simeq (0.3 - 0.4) \,
\mbox{GeV}^2$. In fact, such approach with use of the predictions
of the CQSM has achieved remarkable phenomenological success for
both of the unpolarized and longitudinally polarized
PDFs \cite{WK99},\cite{Wakamatsu03A}.
In the following, we shall therefore use the exact solution
(\ref{Eq:exactRGE}) of the NLO evolution equation with the starting
energy $Q_{ini}^2 = 0.34 \,\mbox{GeV}^2$ to estimate the tensor
charges at a desired scale from the predictions of low
energy models.

\begin{figure}[htb] \centering
\begin{center}
 \includegraphics[width=14.0cm]{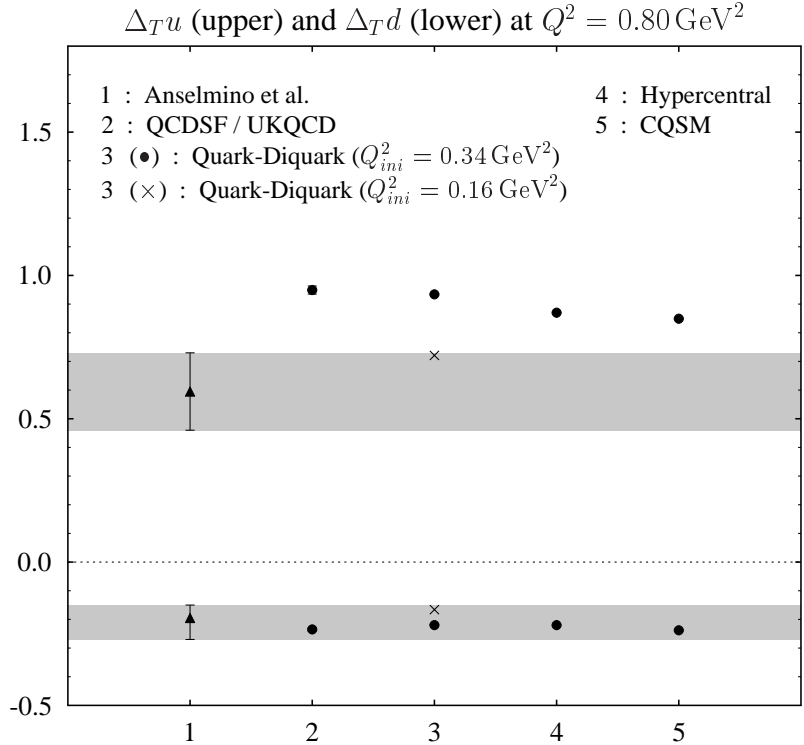}
\end{center}
\vspace*{-1.0cm}
\caption{Comparison of empirical and theoretical tensor charges
for the $u$- and $d$-quarks.
The 1st column and the shaded band stand for
the recent empirical determination of the $u$- and $d$-quark
tensor charges by Anselmino et al., corresponding to the
renormalization scale $Q^2 = 0.8 \,\mbox{GeV}^2$.
The theoretical predictions shown in
other columns are all transformed to the same renormalization
scale with use of the NLO evolution equation (\ref{Eq:exactRGE}).
See the text, for more detail.}
\label{Fig:tchgud}
\end{figure}%

In Fig.\ref{Fig:tchgud}, we compare the 1st empirical information
on the tensor charges for the $u$- and $d$-quarks at the
renormalization scale $Q^2 = 0.8 \,\mbox{GeV}^2$
obtained by Anselmino et al. with the predictions of some 
low energy models as well as that of the lattice QCD.
They all correspond to the scale $Q^2 = 0.8 \,\mbox{GeV}^2$.
For all the low energy models except for the covariant
quark-diquark model of \cite{CBT08}, the starting energy of the
evolution was taken to be $Q_{ini}^2 = 0.34 \,\mbox{GeV}^2$ 
following the discussion above. In the case of the covariant 
quark-diquark model, we have tried two choices of the starting 
energy, i.e. $Q_{ini}^2 = 0.16 \,\mbox{GeV}^2$ and 
$Q_{ini}^2 = 0.34 \,\mbox{GeV}^2$. On the other hand,
the predictions of the lattice QCD are given
in \cite{GHHPRSSZ05} as 
\[
 \Delta_T u \ = \ 0.857 \pm 0.013, \ \ \ \ 
 \Delta_T d \ = \ - \,0.212 \pm 0.005 ,
\]
or 
\[
 \Delta_T \,q^{(I - 1)} \ = \ 1.069 \pm 0.018, \ \ \ \ 
 \Delta_T \,q^{(I = 0)} \ = \ 0.645 \pm 0.018 .
\]
Since these predictions correspond to the renormalization scale 
$Q^2 = 4 \,\mbox{GeV}^2$, we evolve those down by using 
eq.(\ref{Eq:exactRGE}) to obtain the corresponding values 
at $Q^2 = 0.8 \,\mbox{GeV}^2$. One sees that all the theoretical 
predictions for the $d$-quark tensor charge are not largely
different and lie within 
the allowed range of phenomenological extraction.
On the other hand, almost all the theoretical predictions for 
$\Delta_T u$ are larger in magnitude than the empirical 
one, thereby running off the allowed range of the empirical
extraction. The prediction of the covariant quark-diquark model
with use of the starting energy $Q_{ini}^2 = 0.16 \,\mbox{GeV}^2$
is an exception. 
However, we have already pointed out a serious problem
of using such a low starting energy.

At any rate, since the choice of the starting energy for low energy models is rather arbitrary, one must be very careful when making a comparison between model predictions for the tensor charges (or more generally transversity distribution) with phenomenologically
extracted ones. (This should be contrasted with the case of axial
charges. As is widely known, the isovector axial charge is known
 to be scale independent as a
consequence of current conservation.
The isoscalar or flavor-singlet axial 
charge is generally scale dependent, for example, in the standard 
$\overline{\rm MS}$ factorization scheme, because of the $U_A (1)$
anomaly of QCD \cite{AR88}\nocite{CCM88}-\cite{ET88}.
However, this scale dependence is known to be fairly weak except very
low energy.)
Fortunately, we can avoid this troublesome problem of initial
scale choice. The key point is that, since the gluon does not 
couple to the chiral-odd transversities, the evolutions of 
tensor charges are flavor independent.
This in turn means that the {\it ratio} of two tensor charges as 
$\Delta_T d / \Delta_T u$ or 
$\Delta_T q^{(I = 0)} / \Delta_T q^{(I = 1)}$ is totally 
{\it scale independent}.

\begin{figure}[htb] \centering
\begin{center}
 \includegraphics[width=13.0cm,height=10.0cm]{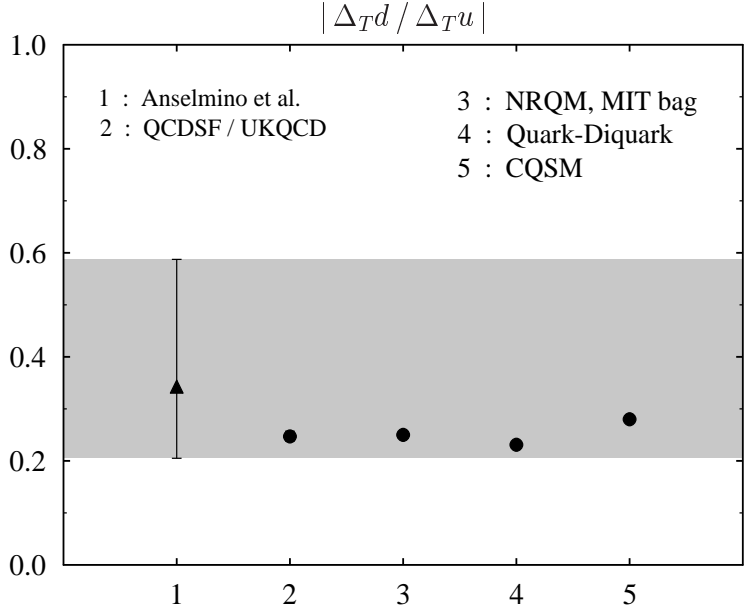}
\end{center}
\vspace*{-1.0cm}
\caption{Comparison of empirical and theoretical tensor charge
ratio $\Delta_T d / \Delta_T u$, which is scale independent.}
\label{Fig:ratio_du}
\end{figure}%

Shown in Fig.\ref{Fig:ratio_du} is the empirically 
extracted tensor-charge ratio $| \Delta _T d / \Delta_T u |$ by
Anselmino et al. in comparison with several theoretical predictions,
i.e. those of lattice QCD \cite{GHHPRSSZ05},
non relativistic quark model (NRQM) or the MIT bag model,
covariant quark-diquark model \cite{CBT08}, and
the CQSM \cite{Wakamatsu07}.
We recall that this ratio is precisely $1/4$ for both of the NRQM
and the MIT bag model. One can convince that
the predictions of all the models as well as that 
of the lattice QCD are not extremely far from this reference value, 
although the prediction of the CQSM is smallest of all.
Since the empirical uncertainties for this ratio is still fairly 
large, we can say that all the theoretical predictions lie within
the error-bars.

\begin{figure}[htb] \centering
\begin{center}
 \includegraphics[width=13.0cm,height=10.0cm]{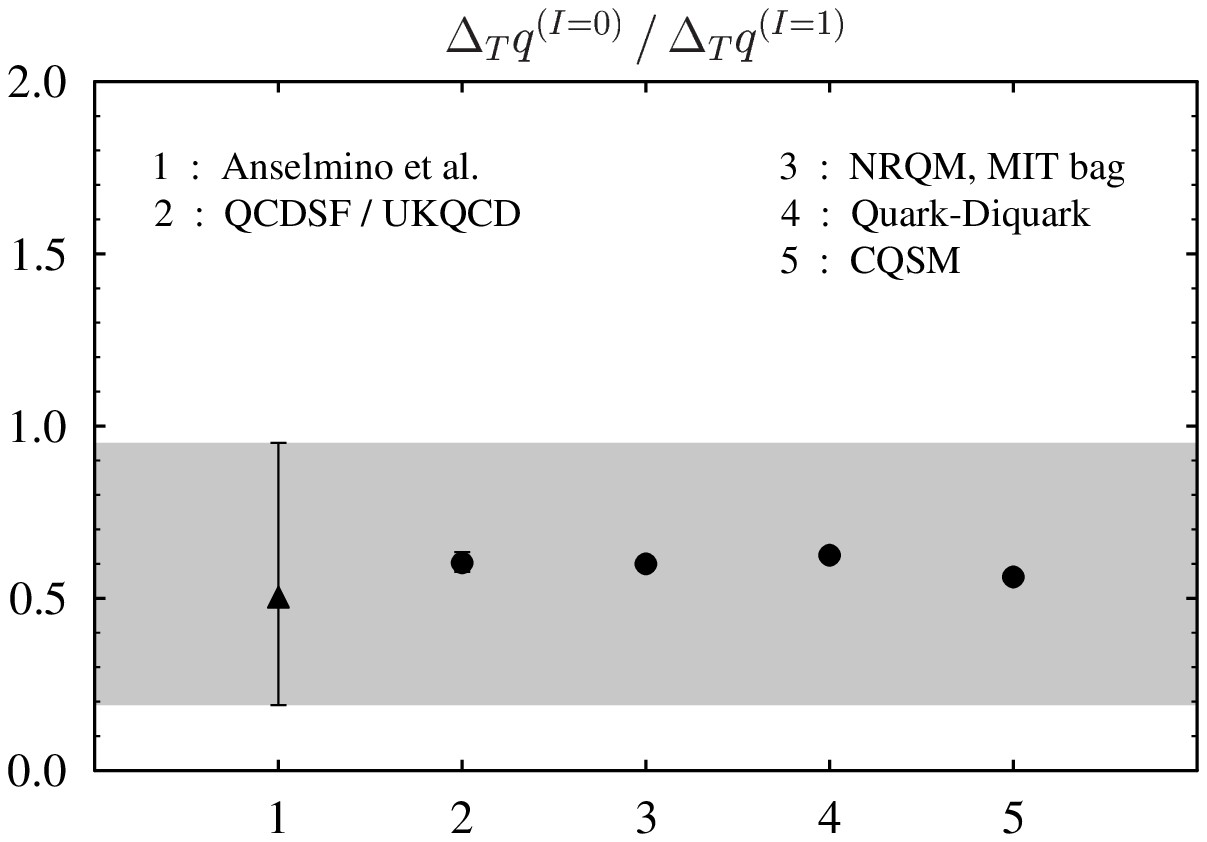}
\end{center}
\vspace*{-1.0cm}
\caption{Comparison of empirical and theoretical tensor charge
ratio $\Delta_T q^{(I=0)} / \Delta_T q^{(I=1)}$, which is scale
independent.}
\label{Fig:ratio_sv}
\end{figure}%

Next, in Fig.\ref{Fig:ratio_sv}, a similar comparison is made
for the tensor 
charge ratio $\Delta_T q^{(I - 0)} / \Delta_T q^{(I = 1)}$.
Again, the prediction of the CQSM gives the smallest value among
all the theoretical predictions. Within the large error-bars,
however, all the theoretical predictions are consistent with 
the phenomenological value.
We emphasize once again that the tensor charge ratio as 
$|\Delta_T d / \Delta_T u |$ and 
$\Delta_T q^{(I = 0)} / \Delta_T q^{(I = 1)}$ are exactly
{\it scale independent} so that it offers a safe and sound basis
of comparison between theoretical predictions and the empirical
extractions. 
Further effort to reduce the uncertainties of phenomenological
extraction would be highly desirable.

\begin{figure}[htb] \centering
\begin{center}
 \includegraphics[width=13.0cm,height=10.0cm]{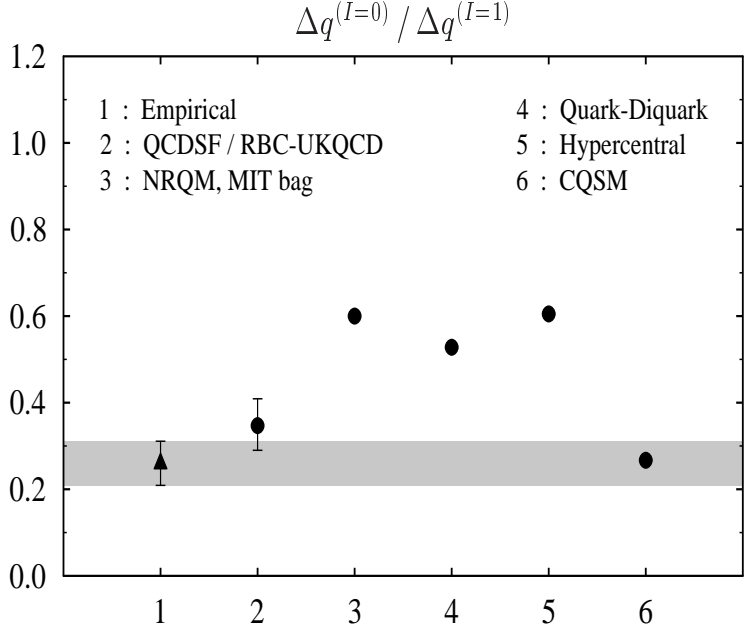}
\end{center}
\vspace*{-1.0cm}
\caption{Comparison of empirical and theoretical axial charge
ratio $\Delta q^{(I=0)} / \Delta q^{(I=1)}$, which is approximately
scale independent.}
\label{Fig:ratio_ga}
\end{figure}

As a general trend, one observes that the predictions for the
tensor charge ratio $\Delta_T q^{(I=0)} / \Delta_T q^{(I=1)}$
by all the low energy models as well as by the lattice QCD are
not extremely far from the reference value of the SU(6) quark model,
i.e. 3/5.
This feature of the tensor charges should be contrasted with that
of axial charges. In Fig.\ref{Fig:ratio_ga}, we compare the
empirically known axial-charge ratio
$\Delta q^{(I=0)} / \Delta q^{(I=1)}$
with the predictions of several models and with that of lattice
QCD. The empirical value here is taken from the HERMES analysis
of the longitudinally polarized structure functions of the
deuteron and proton \cite{HERMES07}. (See also a similar analysis
by COMPASS group \cite{COMPASS05},\cite{COMPASS07}.) 
One sees that fairly small empirical
ratio, which is connected with the famous ``nucleon spin crisis'',
is reproduced only by the CQSM and the lattice QCD, while
the predictions of other low energy models are more or less
close to that of the SU(6) quark model, i.e.
$\Delta^{(I=0)} / \Delta^{(I-1)} = 3 / 5$, thereby largely
overestimating this ratio.  

In any case, as we have repeatedly emphasized,
the possible difference between the axial and
tensor charges of the nucleon (or more generally, the difference
between the longitudinally polarized distribution functions and
the transversity distribution functions of the nucleon) offers
one of the key information for disentangling the internal spin
structure of the nucleon.
Particularly useful here, we think, is the
comparison between the two ratios, i.e.
$\Delta_T q^{(I=0)} / \Delta_T q^{(I=1)}$ and $\Delta q^{(I=0)} / 
\Delta q^{(I=1)}$. As we have emphasized, the former ratio is
exactly scale independent, while the latter has only a weak scale
dependence, so that it offers a safe and sound basis of comparison
between theoretical predictions and empirical extractions for them.
Further effort to reduce the
uncertainties of the phenomenological extraction of the tensor
charges is highly desirable to get more definite information
on the possible difference of these two fundamental quantities.
Do we expect a {\it spin crisis} also for the transverse spins,
or the tensor charges ?

\section{Conclusion \label{Sect:conclusion}}

In summary, we have investigated the forward limit of a particular
combination of chiral-odd generalized parton distributions,
i.e. $G_T (x,0,0) \equiv \lim_{\xi \rightarrow 0, t \rightarrow 0} \,
[\,H_T (x,\xi,t) + 2 \,\tilde{H}_T (x,\xi,t)
+ E_T (x,\xi,t) \,]$ as well as their lower moments within the
framework of the chiral quark soliton model, with particular emphasis
upon the transversity decomposition of the nucleon angular momentum
proposed by Burkardt. 
We found rather strong chiral enhancement near $x \sim 0$ for both
of the isoscalar and isovector GPDs, which reminds us of a similar
chiral enhancement observed in the CQSM predictions for more
familiar unpolarized distribution functions of isoscalar and 
isovector types. We have shown that the $G_T$ is an isoscalar-dominant
quantity, while the isovector component also arises as an $1 / N_c$
correction. In particular, from the 1st moment sum rule of
$G_T$ and $H_T$, we have confirmed a isoscalar dominance of the
``anomalous tensor moments'' $\kappa_T$, which indicates that
the Boer-Mulders functions for the $u$- and $d$-quarks would have
roughly equal magnitude with the same
sign. It should be contrasted with the probable isovector dominance of
the Sivers functions, or of the anomalous magnetic moment of the nucleon.
It is therefore a very important experimental challenge to determine the
relative sign and the magnitudes of the $u$- and $d$-quark
Boer-Mulders functions.

We have also discussed a delicate problem, which may arise
when we try to compare the phenomenologically extracted
tensor charges with corresponding theoretical predictions.
We emphasize that the tensor charges are strongly scale dependent
quantities but the ratios as $\Delta_T d / \Delta_T u$ and 
$\Delta_T q^{(I = 0)} / \Delta_T q^{(I = 1)}$ are exactly scale
independent, so that these ratios are expected to provide us with
a safe and convenient basis of comparison between empirically
determined tensor charges of the nucleon and corresponding
theoretical predictions.


\vspace{3mm}
\appendix

\section{On the 1st moment of $\tilde{E}_T (x,0,0)$}

In this appendix, we shall explicitly verify that the 1st moment of
$\tilde{E}_T (x,0,0)$ vanishes.
At the $O (\Omega^0)$ level, only the isovector part of
$\tilde{E}_T (x,0,0)$ survives as given by eq.(\ref{Eq:ETiv1}).
Its 1st  moment can easily be written down as
\begin{eqnarray}
 \int_{-1}^1 \,\frac{1}{2 \,M_N} \,\,\tilde{E}_T^{(I=1)} (x,0,0) \,dx
 &=& - \,\frac{N_c}{3} \,\sum_{n \in occ} \,
 \langle n \,| \,\tau_3 \,\,i \,x_1 \,
 (\gamma^1 \,\gamma_5 - i \,\gamma^2) \,
 |\, n \rangle \nonumber \\
 &=& - \,\frac{N_c}{18} \,\sum_{n \in occ} \,
 \langle n \,| \, \mbox{\boldmath $\tau$} \cdot
 (\mbox{\boldmath $x$} \times \mbox{\boldmath $\gamma$})\,
 |\, n \rangle .
\end{eqnarray}
Very interestingly, this expression resembles that of the $O (\Omega^0)$
contribution to the isovector magnetic moment of the nucleon in the CQSM
given as
\begin{eqnarray}
 \mu^{(I=1)} (\Omega^0)
 &=& - \,M_N \,\frac{N_c}{9} \,\sum_{n \in occ} \,
 \langle n \,| \, \mbox{\boldmath $\tau$} \cdot
 (\mbox{\boldmath $x$} \times \gamma^0 \,\mbox{\boldmath $\gamma$})\,
 |\, n \rangle ,
\end{eqnarray}
where use has been made of the generalized spherical symmetry of
the hedgehog configuration.
Incidentally, we already know that the time reversal invariance enforces
the 1st moment of $\tilde{E}_T$ to vanish.
As a consistency check of our theoretical framework,
we shall verify it explicitly in the following.
The formal proof in the CQSM utilizes the invariance under the
$G_5$ transformation, which is a simultaneous operations of the
standard time reversal and a flavor SU(2) rotation.
In a standard representation, it is given as
\begin{equation}
 G_5 \ = \ \gamma^1 \,\gamma^2 \,\tau_2,
\end{equation}
and satisfies the following identities
\begin{eqnarray}
 G_5 \,\gamma^\mu \,G_5^{-1} &=& \left( \gamma^\mu \right)^T, \ \ \ \ \ 
 G_5 \,\tau_a \,G_5^{-1} \ = \ - \,\left( \tau_a \right)^T, \\
 G_5 \,\Phi_n (\mbox{\boldmath $x$}) &=& \Phi_n^* (\mbox{\boldmath $x$}),
 \ \ \ \ \ 
 G_5 \,H \,G_5^{-1} \ = \ H^T.
\end{eqnarray}
Using these properties, it is easy to verify the relations
\begin{eqnarray}
 \langle n \,| \, \mbox{\boldmath $\tau$} \cdot
 (\mbox{\boldmath $x$} \times \mbox{\boldmath $\gamma$}) \,|\, n \rangle
 &=& + \,
 \langle n \,| \, \mbox{\boldmath $\tau$} \cdot
 (\mbox{\boldmath $x$} \times \mbox{\boldmath $\gamma$}) \,|\, n \rangle ,
 \nonumber \\
 \langle n \,| \, \mbox{\boldmath $\tau$} \cdot
 (\mbox{\boldmath $x$} \times \gamma^0 \,\mbox{\boldmath $\gamma$})
 \,|\, n \rangle
 &=& - \,
 \langle n \,| \, \mbox{\boldmath $\tau$} \cdot
 (\mbox{\boldmath $x$} \times \gamma^0 \,\mbox{\boldmath $\gamma$})
 \,|\, n \rangle ,
\end{eqnarray}
where use has been made of the reality of the relevant matrix
elements.
These relations then dictates that the 1st moment of $\tilde{E}_T^{(I-1)}$
must vanish identically, while $\mu^{(I=1)}$ need not, as expected.
At the $O (\Omega^1)$ level, both the isoscalar and the isovector parts of
$\tilde{E}_T (x,0,0)$ survive at the first glance. In fact, their
contribution to the 1st moments take the following forms :
\begin{eqnarray}
 &\,& \int_{-1}^1 \,\frac{1}{2 \,M_N} \,\tilde{E}_T^{(I=0)}(x,0,0) \,dx
 \nonumber \\
 &\,& \hspace{0mm} = \ 
 \frac{1}{2 \,I} \,\left( \frac{N_c}{2} \right) \,
 \sum_{m \in nocc, n \in occ} \,\frac{1}{E_m - E_n} \,
 \langle n \,| \,(\mbox{\boldmath $x$} \times \mbox{\boldmath $\gamma$})_3
 \,| \,m \rangle \,\langle m \,| \,\tau_3 \,| \,n \rangle, \\
 &\,& \int_{-1}^1 \,\frac{1}{2 \,M_N} \,\tilde{E}_T^{(I=1)}(x,0,0) \,dx
 \nonumber \\
 &\,& \hspace{0mm} = \ 
 i \,\varepsilon_{3 a c} \,\,\frac{N_c}{6 \,I} \,
 \sum_{m \in nocc, n \in occ} \,\frac{1}{E_m - E_n} \,
 \langle n \,| \,\tau_a \,x_1 \,\gamma_2
 \,| \,m \rangle \,\langle m \,| \,\tau_c \,| \,n \rangle.
\end{eqnarray}
However, it is not so difficult to prove that both of the above
expressions vanishes owing to the symmetry under the $G_5$
transformation. Since the SU(2) isospin symmetry is naturally
respected in our effective theory, this just reconfirms
the general statement that the 1st moments
$\tilde{E}_T^{(I=0,1)}$ vanish by time reversal invariance.


\end{document}